\documentclass[aps,prl,twocolumn,superscriptaddress]{revtex4-1}

\usepackage{graphicx}
\usepackage{dcolumn}
\usepackage{bm}
\usepackage{natbib}
\usepackage{color}
\usepackage{graphics}
\usepackage{pgf}
\usepackage{multimedia}
\usepackage{mathptmx}
\usepackage{mathrsfs}
\usepackage{amsfonts}
\usepackage{pdfpages}

\begin{document}

\title{Suppression of the ferromagnetic order in the Heusler alloy Ni$_{50}$Mn$_{35}$In$_{15}$ by hydrostatic pressure}

\author{C. Salazar Mej\'{i}a}
\email{Catalina.Salazar@cpfs.mpg.de}
\affiliation{Max Planck Institute for Chemical Physics of Solids, N\"{o}thnitzer Str.\ 40, 01187 Dresden, Germany.}
\author{K. Mydeen}
\affiliation{Max Planck Institute for Chemical Physics of Solids, N\"{o}thnitzer Str.\ 40, 01187 Dresden, Germany.}
\author{P. Naumov}
\affiliation{Max Planck Institute for Chemical Physics of Solids, N\"{o}thnitzer Str.\ 40, 01187 Dresden, Germany.}
\author{S.~A. Medvedev}
\affiliation{Max Planck Institute for Chemical Physics of Solids, N\"{o}thnitzer Str.\ 40, 01187 Dresden, Germany.}
\author{C. Wang}
\affiliation{Max Planck Institute for Chemical Physics of Solids, N\"{o}thnitzer Str.\ 40, 01187 Dresden, Germany.}
\author{M. Hanfland}
\affiliation{ESRF, BP220, 38043 Grenoble, France.}
\author{A.~K. Nayak}
\affiliation{Max Planck Institute for Chemical Physics of Solids, N\"{o}thnitzer Str.\ 40, 01187 Dresden, Germany.}
\affiliation{Max Planck Institute of Microstructure Physics, Weinberg 2, 06120 Halle, Germany.}
\author{U. Schwarz}
\affiliation{Max Planck Institute for Chemical Physics of Solids, N\"{o}thnitzer Str.\ 40, 01187 Dresden, Germany.}
\author{C. Felser}
\affiliation{Max Planck Institute for Chemical Physics of Solids, N\"{o}thnitzer Str.\ 40, 01187 Dresden, Germany.}
\author{M. Nicklas}
\email{nicklas@cpfs.mpg.de}
\affiliation{Max Planck Institute for Chemical Physics of Solids, N\"{o}thnitzer Str.\ 40, 01187 Dresden, Germany.}

\date{\today}

\begin{abstract}

We report the effect of hydrostatic pressure on the magnetic and structural properties of the shape-memory Heusler alloy Ni$_{50}$Mn$_{35}$In$_{15}$. Magnetization and x-ray diffraction experiments were performed at hydrostatic pressures up to 5~GPa using diamond anvil cells. Pressure stabilizes the martensitic phase, shifting the martensitic transition to higher temperatures and suppresses the ferromagnetic austenitic phase. Above $\sim3$~GPa, where the martensitic-transition temperature approaches the Curie temperature in the austenite, the magnetization shows no indication of ferromagnetic ordering anymore. We further find an extremely large temperature region with a mixture of martensite and austenite phases, which directly relates to the magnetic properties.

\end{abstract}

\maketitle

Heusler alloys which exhibit a martensitic structural transformation in proximity to a ferromagnetic (FM) phase have attracted much attention due to the multiple functional properties connected to the coupling of the structural transition to magnetic degrees of freedom, such as shape memory \cite{Kainuma2006,Planes_JOPM_2009,Krenke_PRB_2007}, magnetocaloric \cite{Liu2012,GhorbaniZavareh2014}, and barocaloric effects \cite{Manosa2010}.
In the austenitic phase in NiMn-based alloys the Mn moments order ferromagnetically, which arises mainly due to the RKKY-exchange interaction \cite{Buchelnikov2008,Sasioglu2008,RamaRao2014}. In the martensitic state, which can form in a simple tetragonal, a complex monoclinic, or an orthorhombic layered structure, a strong competition between FM and antiferromagnetic interactions exists, leading to a high sensitivity of the physical properties on the interatomic distances. The application of pressure is, therefore, an important tool to study the relationship of magnetism and crystal structure, without altering the intrinsic properties unintentionally, or introducing additional disorder in the structure, like in the case of element substitution. In Ni-Mn-$Z$ ($Z={\rm In}$, Sb, Sn), application of a small pressure $p\lesssim1$~GPa stabilizes the martensitic phase and, therefore, the martensitic transition temperature increases strongly upon increasing pressure, while the effect on the Curie temperature in the austenitic phase, $T_C^A$, is rather small \cite{Aksoy2007,Manosa2008,Nayak2009,Sharma2011}. In closely related compounds, it has been reported that low pressures can improve the magnetocaloric effect \cite{Sharma2011} or lead to a large barocaloric effect \cite{Manosa2010}.

At ambient pressure, the shape-memory Heusler alloy Ni$_{50}$Mn$_{35}$In$_{15}$ undergoes on cooling a paramagnetic to FM transition at $T_C^A\approx313$~K, followed by a first-order martensitic structural transformation from a cubic high-temperature to a low-temperature modulated structure \cite{Yan2015} at $T_M\approx248$~K \cite{Nayak2014,GhorbaniZavareh2014}. On heating the reverse martensitic transition takes place at $T_A\approx261$~K. The magnetostructural transition drives the material from the FM state to a state with a small remaining magnetization. Upon further cooling ferrimagnetic order develops in the martensitic phase below $T_C^M\approx200$~K \cite{Nayak2014,GhorbaniZavareh2014}.

In this Letter, we study the effect of hydrostatic pressure on the magnetic and structural properties of the Heusler alloy Ni$_{50}$Mn$_{35}$In$_{15}$. While the FM transition in the austenite at $T_C^A$ displays only a weak pressure dependence, the martensitic transition temperature $T_{A,M}$ increases strongly upon increasing pressure. This leads to a suppression of the FM phase. Our structural investigation indicates a large mixed-phase region of austenite and martensite phases extending to temperatures far away from the martensitic transformation. This result is consistent with the FM ordering being restricted to the austenitic phase and evidences the strong interrelation of structural and magnetic properties in Ni-Mn-In shape-memory Heusler alloys.

Polycrystalline ingots of Ni$_{50}$Mn$_{35}$In$_{15}$ were prepared as previously reported \cite{Nayak2014}. A part of the sample was crushed in small pieces for magnetization measurements. For the x-ray diffraction (XRD) experiments powder was prepared by milling some material down to a grain size smaller than 20~$\mu$m. To reduce the residual mechanical stresses in the grains, the powder was annealed at 800$^\circ$C for 4~h under argon atmosphere. The temperature dependence of the magnetization was recorded at pressures up to 4.6~GPa using a miniature diamond anvil cell (DAC) in a magnetic property measurement system (Quantum Design). In this experiments glycerin served as pressure-transmitting medium. Powder XRD at ambient pressure was performed at the National Synchrotron Radiation Research Center (NSRRC, Taiwan) and under applied pressure at the European Synchrotron Radiation Facility (ESRF, France) at the beamline ID09 up to a maximum pressure of 5~GPa using a DAC with neon as a pressure-transmitting medium. For thermalizing the sample a liquid helium cooled cryostat and an external resistive heating device were used.
The pressure inside the DACs was determined by a standard ruby fluorescence method.

\begin{figure}[t!b]
\begin{center}
  \includegraphics[width=0.9\columnwidth]{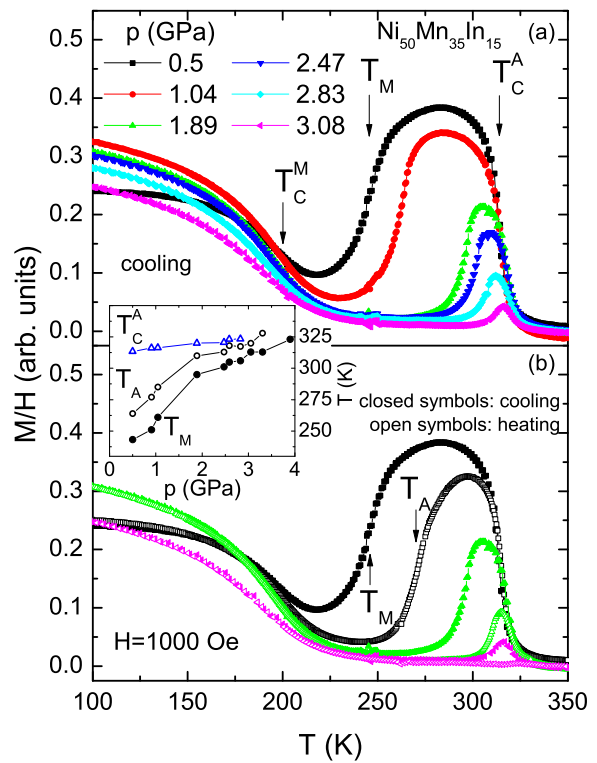}
  \caption{Susceptibility data for selected hydrostatic pressures recorded on cooling (solid symbols) and heating (open symbols). The transition temperatures are marked only for $p=0.5$~GPa. The inset displays the evolution of $T_C^A$, $T_M$, and $T_A$ with pressure.}\label{MvsT4}
\end{center}
\end{figure}

Figure \ref{MvsT4} shows  selected magnetization curves as function of temperature for different pressures. For each measurement, the desired pressure was applied at room temperature (RT). Afterwards, the sample was heated up to 350~K where a magnetic field of 1000~Oe was applied and the magnetization was recorded upon cooling down to 10~K followed by a heating cycle up to 350~K again. We carried out two pressure experiments with maximum pressures of 3.08 and 4.6~GPa, respectively. We note that we cannot provide absolute values of the magnetization due to the large uncertainty in the determination of the sample mass. However, the relative changes between different pressures in one pressure experiment are not affected by this and reflect, therefore, pressure-induced changes in the sample magnetization.

The magnetization curves recorded at the lowest pressure of 0.5~GPa (see Fig.\ \ref{MvsT4}) display the same characteristics than the data previously reported at ambient pressure taken on a sample from the same batch \cite{Nayak2014,GhorbaniZavareh2014,supp}, i.e., upon cooling, the FM phase transition in the austenite at $T_C^A$, the martensitic transition at $T_M$ from a FM cubic austenite to a martensite displaying a strongly reduced magnetization and at lower temperatures the ferrimagnetic transition at $T_C^M$. The transition temperatures are defined by the corresponding inflection points in the $M(T)$ curves. Increasing pressure causes only a weak increase in $T_C^A(p)$ consistent with reports in literature for other Ni-Mn based Heusler alloys \cite{Kanomata1987}. The more pronounced effect is observed on $T_{A,M}$, indicated by the drop in magnetization toward lower temperatures at $T_M$ (at $T_A$ upon increasing temperature), which increases strongly upon increasing pressure. Thus, the application of pressure stabilizes the martensitic phase. The effect of pressure on $T_C^A$ is much weaker than that on $T_{A,M}$. Thus, upon increasing pressure $T_{A,M}(p)$ approaches $T_C^A(p)$ as exemplified in the $T-p$ phase diagram depicted in the inset of Fig.\ \ref{MvsT4}. The distance between the two transitions $T_C^A-T_{M}$ decreases until the $T_{A,M}(p)$ phase line crosses the $T_C^A(p)$ phase line and no indication of a magnetic phase transition remains in the magnetization data. The suppression of the FM order is accompanied by a strong reduction of the magnetization in the FM phase upon increasing pressure.

Now we turn to the thermal hysteresis observed in the magnetization curves. Hysteretic effects are not only restricted to the immediate vicinity of the first-order martensitic phase transition, but extend to much higher and also lower temperatures. At low pressures, the data for 0.5~GPa are shown in Fig.\ \ref{MvsT4}b, the hysteresis opens already below $T_C^A$ and closes just before $T_C^M$. This hints at a large coexistence region of austenite and martensite phases, as will be shown below.  We further observe a finite magnetization between $T_C^M$ and $T_{A,M}$. We propose that this magnetization is not reflecting any ordering in the martensitic phase, but is instead caused by weakly coupled ferromagnetically ordered austenitic regions still present in a non-magnetic martensitic background. Upon further increasing pressure $T_{A,M}(p)$ increases, but $T_C^M(p)$ remains almost unchanged. Therefore, the distance between the two transitions increases and we observe a closing of the hysteresis in $M(T)$ well above $T_C^M$. We note that we do not observe any substantial magnetization in the region between $T_{A,M}$ and $T_C^M$ once the hysteresis is closed. We take this finding as evidence that the FM ordering is restricted to the austenitic phase. We cannot exclude antiferromagnetic or spin-glass type of order in the martensitic phase from our data. Moreover, the maximum value of $M(T)$ reached in the FM phase differs strongly between cooling and heating cycles. While there is almost no difference at ambient pressure \cite{Nayak2014,GhorbaniZavareh2014,supp}, the difference grows with increasing pressure. At 3.08~GPa, we only observe a small kink in the cooling curve, but almost no anomaly related to the FM transition on heating is visible anymore (see Fig.\ \ref{MvsT4}b). Above this pressure it is difficult to detect any signature of the transition in either cooling or heating curves. At 4.6~GPa, the highest pressure in our experiment, no apparent transition anomaly is visible anymore (not shown).

The $T-p$ phase diagram of Ni$_{50}$Mn$_{35}$In$_{15}$ determined from the magnetization data is depicted in the inset of Fig.\ \ref{MvsT4}. The pressure evolution of $T_{A,M}$ can be divided in two regions, one at low pressure $p\lesssim1.9$~GPa with $dT_{A(M)}(p)/dp\approx33$~K/GPa (36~K/GPa) and one for $p\gtrsim1.9$~GPa with $dT_{A(M)}(p)/dp\approx12$~K/GPa (14~K/GPa). $T_C^A(p)$ exhibits only a weak almost linear pressure dependence with a slope of  $dT_C^A(p)/dp\approx 2.3$~K/GPa. We note that the thermal hysteresis between $T_M$ and $T_A$ also decreases above $\sim1.9$~GPa.

In order to relate the pressure evolution of the magnetic properties of Ni$_{50}$Mn$_{35}$In$_{15}$ with the structural changes XRD experiments were carried out. We note that the grinding process, required for producing the powder needed for the experiments, induced stresses in the grains which could not be completely removed by a second heat treatment. As a result $T_{A,M}$ is shifted by approximately 24~K toward higher temperatures at ambient pressure in the powdered material compared with the bulk sample (see the Supplemental Material for details \cite{supp}). However, the residual stresses in the powder do not affect the general characteristics of the material.

\begin{figure}[t!b]
\begin{center}
  \includegraphics[width=0.9\columnwidth]{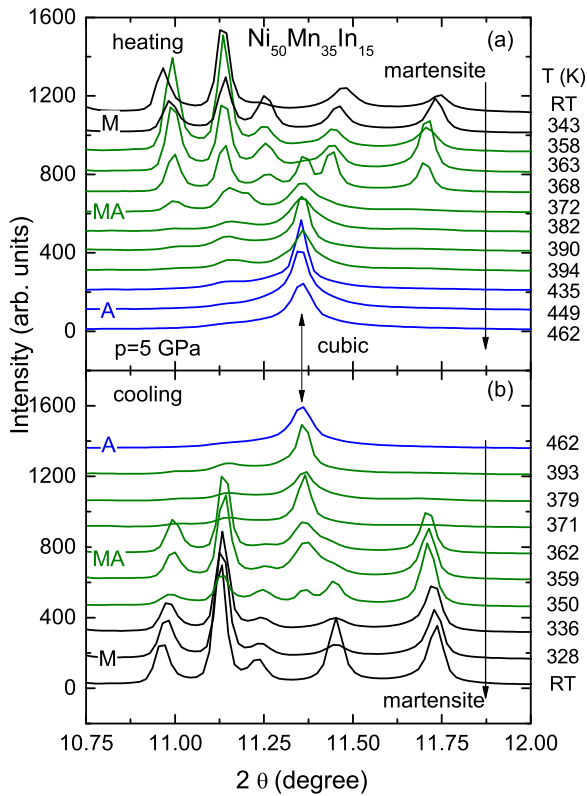}
  \caption{Diffraction patterns collected with $\lambda=0.415024$~{\AA} at different temperatures taken during (a) heating and during (b) cooling at 5~GPa. The austenitic phase is labeled by the letter A (blue), the mixed martensitic and austenitic phases by MA (green), and the martensitic phase by M (black).}\label{p5all}
\end{center}
\end{figure}

Ambient pressure XRD data confirm a cubic austenitic phase with lattice parameter $a=6.00509(7)$~\AA\ at high temperatures, while the low-temperature martensitic phase is a complex modulated structure \cite{supp,Yan2015}. The XRD experiments under pressure were performed between 1.5 and 5~GPa. The pressure was always changed at RT and diffractograms were taken during cooling and heating cycles. The end temperatures were chosen in order to obtain a single phase material. Figure \ref{p5all} displays selected diffractograms recorded at 5~GPa during heating from RT up to 462~K and during cooling down to RT again in the range $10.75 ^\circ\leq2\theta \leq12 ^\circ$. In this window only the (220) peak of the cubic austenitic phase is observed, but five peaks corresponding to the martensitic phase. Due to the complexity of the modulated structure we refrain our analysis to the temperature evolution of the martensitic transformation.

We first focus on the diffractograms obtained on the heating cycle depicted in Fig.\ \ref{p5all}a for 5~GPa. At RT and 343~K only the peaks corresponding to the martensitic phase are present indicating a single phase. Upon increasing temperature, the (220) peak of the cubic austenite structure appears at 358~K. The martensitic and the austenitic phase coexist for a large temperature range from 358~K up to 394~K. Above $\sim400$~K, Ni$_{50}$Mn$_{35}$In$_{15}$ transforms completely to the austenitic phase and only the cubic (220) peak is visible. Upon cooling, displayed in Fig.\ \ref{p5all}b, we observe a thermal hysteresis.

The unit-cell volume of the austenitic phase decreases linearly with pressure from $V=216.96$~\AA$^3$ at ambient pressure to $V=208.80$~\AA$^3$ at 5~GPa. The lattice parameters of the cubic phase were determined from the diffractograms at 420~K, in order to have a single phase material (see the Supplemental Material for details \cite{supp}).

\begin{figure}[t!b]
\begin{center}
  \includegraphics[width=0.9\columnwidth]{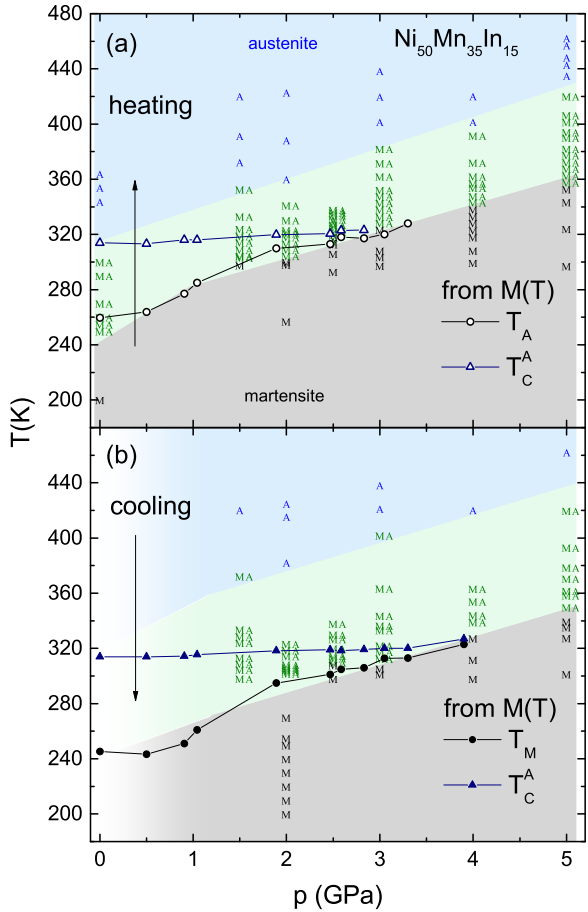}
  \caption{$T-p$ phase diagram of Ni$_{50}$Mn$_{35}$In$_{15}$ determined on (a) heating and (b) cooling cycles by XRD and magnetization experiments. The austenitic phase is labeled by the letter A (blue), the mixed martensitic and austenitic phases by MA (green), and the martensitic phase by M (black).}\label{PhaseDiagramLett3}
\end{center}
\end{figure}

Our results on the phase diagram of Ni$_{50}$Mn$_{35}$In$_{15}$ are summarized in Fig.\ \ref{PhaseDiagramLett3}, in the upper panel for the heating and in the lower panel for the cooling cycle. We note that at ambient pressure no XRD data were recorded on cooling. Due to the large coexistence region of the austenite and martensite it is not possible to infer a martensitic transition temperature from the structural data. The extent of the mixed-phase region does not change upon increasing pressure, but shifts to higher temperatures at about the same rate as $T_{A,M}$ determined from the magnetization data. This evidences the expected strong coupling between structural and magnetic properties at the martensitic transformation. On the other hand, the FM transition temperature exhibits almost no temperature dependence and is independent of the pressure evolution of the mixed-phase region. Consequently, the FM phase disappears once the fraction of the austenitic phase gets too small, i.e., $T_{A,M}\gtrsim T_C^A$, and no long-range order can develop anymore. This strongly suggests that the FM order is bound to the austenitic phase.

The magnetic and structural data confirm that $T_{A,M}(p)$ increases strongly with pressure (see Fig.\ \ref{PhaseDiagramLett3}). This can be understood from thermodynamics considering that pressure stabilizes the phase with smaller unit-cell volume, i.e., the martensitic phase \cite{Manosa2008,Li2009}. According to the Clausius-Clayperon equation, the shift of a first-order phase transition with pressure is given by $dT/dp=\Delta V_M/\Delta S$, where $\Delta V_M$ and $\Delta S$ are the changes in the molar volume and in entropy at the transition, respectively. For Ni$_{50}$Mn$_{35}$In$_{15}$ a relative volume change of $\Delta V/V\approx0.3$\% \cite{Manosa2008} and an entropy change of $\Delta S=10.3$~Jkg$^{-1}$K$^{-1}$ have been reported \cite{SalazarMejia}. Considering $V_M\approx1.30\times 10^{-4}$m$^3$kg$^{-1}$ in the austenite, we obtain $dT/dp \approx38$~K/GPa, which is in good agreement with our experimental result for pressures below 1.9~GPa, $dT/dp \approx36$~K/GPa. For $p\gtrsim1.9$~GPa $dT/dp$ is considerably smaller. This is most likely caused by an increase of the entropy change at the martensitic transition, due to a reduction of the magnetic contribution to the Gibbs free energy, since the distance between $T_C^A$ and $T_{A,M}$ decreases \cite{SalazarMejia,Ito2007}. The magnetic entropy change related with the change in magnetization at the martensitic transition has an opposite sign compared to that of the entropy change related with the structural transition. Therefore, upon increasing pressure the net entropy change at the transition increases since the  structural contribution is supposed to be pressure independent. At 3~GPa, the molar volume of the austenitic phase is $V_M\approx1.27\times 10^{-4}$m$^3$kg$^{-1}$ ($a=5.9588(3)$\,\AA). Assuming a constant relative volume change at the transition and considering the experimental value $dT_M/d_p\approx 14$~K/GPa, the entropy change at the martensitic transition increases to $\Delta S\approx27$~Jkg$^{-1}$K$^{-1}$.

With increasing pressure $T_{A,M}(p)$ moves closer to the Curie temperature $T_C^A$ and the change in magnetization at the FM transition decreases. This can be understood in the following way (see Fig.\ \ref{PhaseDiagramLett3}b): at ambient pressure around 330~K the sample is in the fully austenitic state. Upon lowering the temperature the whole sample orders ferromagnetically at $T_C^A$ reaching a magnetization value corresponding to the full sample volume. Upon further cooling the martensitic transition takes place and the magnetization drops, due to the different magnetic properties of the austenitic and the martensitic phase. At applied pressure, for instance at $2$~GPa, at 330~K the sample consists of a mixture of martensitic and austenitic parts. Upon cooling only the moments in the austenitic phase order ferromagnetically. Accordingly, the measured magnetization is reduced corresponding to the fraction of the austenitic phase in the sample. As a consequence the magnetization change at $T_M$ decreases too. At $3$~GPa almost all of the sample has already transformed to the martensitic phase at $T_C^A$. Thus, only the remaining austenitic phase orders ferromagnetically at $T_C^A$ leading to a tiny change in the magnetization as can be seen in Fig.\ \ref{MvsT4}. Once $T_M(p)$ becomes larger than $T_C^A(p)$ no long range FM ordering is observed anymore.

The mixed-phase region of austenite and martensite phases is not restricted to the immediate vicinity of the thermal hysteresis region of the martensitic transition as determined by the magnetization data. We find that the mixed-phase region shifts linearly to higher temperature upon increasing pressure. This leads, assuming a constant temperature, to a growing fraction of the martensitic phase and a declining contribution of the austenitic phase upon increasing pressure. Since $T_C^A$ is almost pressure independent the fraction of the austenitic phase in the ferromagnetically ordered region decreases. At the same time, we observe a reduction in the size of the magnetization in the FM phase. Therefore, we conclude that the FM order is bound to the austenitic phase and the decrease in the magnetization reflects the decrease in the fraction of the austenitic phase. Following the same arguments we can understand the observation of a relatively large magnetization between $T_C^M$ and $T_{A,M}$ in the predominantly martensitic phase at low pressures. In this regime, we still find a small fraction of the FM austenite present. Upon increasing pressure the mixed-phase region moves to higher temperatures and the fraction of FM austenite decreases further and only the martensitic phase remains. The large mixed-phase region is a critical issue for applications since it has a strong influence on the magnetic properties in a wide temperature region and not only around the martensitic phase transformation.

Finally, we compare the effect of hydrostatic pressure in Ni$_{50}$Mn$_{35}$In$_{15}$ with chemical substitution. Substitution of Mn by In in the series Ni$_{50}$Mn$_{25+x}$In$_{25-x}$ shows a similar result as the application of external pressure: the martensitic transition shifts to higher temperatures and at lower In concentration no FM austenitic phase is present \cite{Krenke2006,Kanomata2009,Yan2013}. Like the application of hydrostatic pressure, a decrease in the In content leads to a reduction in the unit-cell volume, which is attributed to the difference in the ionic radii of the Mn and In atoms \cite{Kanomata2009}. Furthermore, the entropy change at the martensitic transition increases with decreasing In content. In particular, $\Delta S$ is larger for the samples where the martensitic transition takes place between two non-magnetic phases \cite{Krenke2006,Yan2013,Stern-Taulats2015}, in agreement with our results.

In summary, in Ni$_{50}$Mn$_{35}$In$_{15}$ application of hydrostatic pressure suppresses the FM ordering. While $T_C^A$ only shows a weak pressure dependence, $T_{A,M}$ shifts strongly to higher temperatures upon increasing pressure. The latter confirms the expected strong coupling of the magnetic and structural properties at the martensitic transition. The pressure evolution of the magnetic properties can be understood considering the extremely large martensite/austenite mixed-phase region. Our findings show that in Heusler shape-memory alloys even the second order FM phase transition in the austenitic phase can be influenced by the martensitic transformation and the related hysteretic behavior.

\begin{acknowledgments}
This work was financially supported  by the ERC Advanced Grant (291472) "Idea Heusler".
The XRD experiments at ambient pressure were performed at the National Synchrotron Radiation Research Center (NSRRC, Taiwan) and the experiments under pressure were performed at the ID09A beamline at the European Synchrotron-Radiation Facility under proposal HC-1342.
\end{acknowledgments}


\end{document}